\documentstyle[aps,preprint,tighten,floats,psfig,epsf,rotate]{revtex}

\draft
 \preprint{\vbox{\hfill
 \rm \null \hfill ADP-01-20/T455
 }}

\begin{document}

\title{\bf Symanzik Improvement In The Static Quark Potential }

\author{
Fr\'{e}d\'{e}ric
D. R. Bonnet$^1$\footnote{E-mail:~fbonnet@physics.adelaide.edu.au}, 
Derek B. Leinweber$^1$\footnote{E-mail:~dleinweb@physics.adelaide.edu.au},
Anthony G.
Williams$^{1,2}$\footnote{E-mail:~awilliam@physics.adelaide.edu.au},
\\ 
James M. Zanotti$^1$\footnote{E-mail:~jzanotti@physics.adelaide.edu.au}}
\address{$^1$Special Research Center for the Subatomic Structure of
Matter (CSSM) and \\ 
Department of Physics and Mathematical Physics, University of Adelaide
5005, Australia.} 
\address{$^2$Department of Physics and SCRI, Florida State University,
Tallahassee, FL 32306-4052} 
\date{\today}
\maketitle

\begin{center} \bf{Abstract} \end{center} 

A systematic investigation of Symanzik improvement in the gauge field
action is performed for the static quark potential in quenched QCD.  We 
consider Symanzik improved gauge field configurations on a $16^3 \times 32$ 
lattice with a relatively coarse lattice spacing of 0.165(2) fm.  A matched 
set of Standard Wilson gauge configurations is prepared at $\beta = 5.74$ 
with  the same physical volume and lattice spacing and is studied for 
comparison.  We find that, despite the coarse lattice spacing, the unimproved and 
less-expensive Wilson action does as well as the Symanzik action in allowing
us to extract the static quark potential at large $q\bar{q}$ separations.  We
have considered novel methods for stepping off-axis in the 
static quark potential which provides new insights into the extent to which the 
ground state potential dominates the Wilson loop correlation function.

\vspace{3mm}
PACS number(s): 12.38.Aw, 12.38.Gc, 12.39.Pn, 14.70.Dj

\newpage

\section{Introduction}

While string breaking in the static quark potential of full
dynamical-fermion QCD has long been predicted, direct observation of
string breaking in lattice QCD is proving to be elusive
\cite{Kuti:1999,Aoki:1999,Talevi:1999,Bali:1998,Duncan:2000kr,Pennanen:2000yk,Trottier:1999}.  
The main reason for
this appears to be due to difficulty in isolating the ground state of
the static quark potential at large $q\bar{q}$ separations
\cite{Trottier:1999}.  Overlap of standard operators for separating a
static $q\bar{q}$ pair with excited states of the potential and close
spacing in the spectrum of the potential at large separations demand
significant Euclidean time evolution in order to isolate the ground
state \cite{Trottier:1999,Gusken:1998,Drummond:1998}.  

Accessing large Euclidean times is extremely difficult.  APE smearing
\cite{ape} is widely recognized as an effective way to approach this
region.  The smearing of the spatial links of the lattice mocks up the
flux tube joining two static quarks in the ground state potential.
The smeared operator provides better overlap between the vacuum and
the ground state potential and improves the signal to noise ratio in
the correlation function \cite{Legeland:1998rs}.

As one approaches the large Euclidean space-time regime, statistical errors
grow exponentially.  While it is easy to fit the lattice data for the
effective potential to a plateau ansatz, it is difficult to have
confidence that the asymptotic value has been reached.  With exponentially
growing error bars, correlated $\chi^2$ and goodness of fit parameters can
provide a lower bound on the Euclidean time regime, but do not provide
information on whether such a bound is sufficient to isolate the ground
state potential.

Techniques for evaluating the extent to which the ground state dominates the
Wilson loop are needed.  Fortunately for the standard Wilson gluon
action, methods exist.  However, these methods break down when improved 
actions are used.

Recently, novel ideas have been explored in the search for string
breaking
\cite{Trottier:1999,DeTar:1999ie,Knechtli:1999av,Knechtli:1998gf}.  We
are motivated by the encouraging results of Ref. \cite{Trottier:1999}
studying string breaking via Wilson loops in $2+1$ dimensional QCD.
There, using improved actions on coarse lattices in three dimensions,
string breaking was observed.  These authors emphasize that Euclidean
time evolution of the order of 1 fm is required to isolate the ground state
potential.  The efficiency afforded by the use of 
coarse lattices is argued to be key to achieving this goal.

Here we explore a systematic comparison of the static quark potential
obtained from standard Wilson and Symanzik improved field configurations.  
Given the same number of configurations, similar lattice spacing and equivalent
analysis techniques, we illustrate how the Standard Wilson action does as well
as the Symanzik improved action in extracting the long-range $q\bar{q}$ potential.
 While the authors of 
Ref.\ \cite{Trottier:1999} emphasize
the efficiency of the improved action approach, we find that unimproved
actions on reasonably coarse lattices offer the most efficient and suitably
accurate use of limited computer resources.

Section~\ref{SQP} briefly explains our technique for calculating the 
static quark potential from Wilson loops.  Here we present an alternative 
way of exploring the off-axis static quark potential.  This new method provides
additional information on the extent to which the ground-state potential dominates
the Wilson loop and can provide confidence that sufficient evolution in Euclidean 
time has occurred.  Section~\ref{action} 
outlines the Symanzik improved action.  In Section~\ref{simRes}, we present and
discuss our results and conclude in Section~\ref{conclusion}.

\section{The Static Quark Potential}
\label{SQP}

The spectrum of the static quark potential is determined from Wilson
loops $W(r,t)$ of area $r\times t$, 
\begin{equation}
W(r,t) = \sum_i C_i(r)\, \exp(-V_i(r) \, t) \, .
\label{ground} 
\end{equation}
In order to enhance $C_1(r)$, which measures the overlap of the loop with
the ground state potential, the spatial links are APE smeared
\cite{ape}.  This smearing procedure replaces a spatial link
$U_{\mu}(x)$, with a sum of the link and $\alpha$ times its spatial
staples:
\begin{eqnarray}
U_{\mu}(x)\rightarrow (1-\alpha)U_{\mu} 
&+& \frac{\alpha}{4}\sum_{\nu=1 \atop \nu\neq\mu}^{3}\Big[U_{\nu} (x) 
U_{\mu}(x+\nu a)U_{\nu}^{\dag}(x+\mu a)
\nonumber \\ \mbox{} 
&+& U_{\nu}^{\dag}(x-\nu a)U_{\mu}(x-\nu a)U_{\nu}(x-\nu a +\mu a)\Big] \,.
\label{smear}
\end{eqnarray}
This is applied to all spatial links on the lattice followed by
projection back to SU(3), and repeated $n$ times.  

Tuning the smearing parameters is key to the success of the approach.  They
govern both the Euclidean time extent of the Wilson loop correlation function and
the relative contributions of ground to excited state contributions.  It is
vital to have a quantitative technique for tuning these parameters in order to 
optimise future studies of string breaking.  The measure must
be the extent to which the ground state dominates the correlation function.

Efficient methods exist for the unimproved Wilson action for fine tuning the smearing 
parameters to provide optimal overlap with the ground state potential. 
For $t=0$, $W(r,t=0) = 1$ providing the constraint $\sum_i C_i(r) = 1$ for a
given $r$.  For unimproved actions, where the transfer matrix is positive
definite, each $C_i (r) \ge 0$.  This means $C_1(r)$ can be monitored at large $r$
but small $t$
as the number of smearing sweeps are varied, with the optimal amount of smearing
occurring when $C_1(r)\approx 1$.
The proximity of $C_1(r)$ to 1 for small $t$ may be easily
estimated from the ratio
\begin{equation}
W^{t+1}(r,t) / W^t(r,t+1)
\end{equation}
which equals $C_1(r)$ in the limit $C_1(r) \to 1$.  This provides a quantitative measure 
of ground-state-dominance for unimproved Wilson actions.  We note that it is sufficient 
\cite{Bonnet:2000dc} to fix the smearing fraction, $\alpha$, and explore the parameter 
space via the number of smearing sweeps, $n$.

This procedure can be repeated for 
a number of alternate paths of links for a given separation $r$.  By using variational 
techniques as described in Ref.\cite{Morningstar:1997ff}, the combination of paths that gives 
the greatest overlap with the ground state can be found.

Actions improved in the time direction do not satisfy Osterwalder-Schrader \cite{Osterwalder:1973dx}
positivity \cite{Parisi:1985iv}.
This spoils the positive definite nature of the transfer matrix and the constraint
$C_i (r) \ge 0$ is lost.  Hence one needs either a new quantitative measure for evaluating 
ground-state-dominance, or to check the merits for using Symanzik improved gluon actions for the 
static quark potential at large $r$.

For exploratory purposes, we fixed $\alpha=0.7$ and considered the values $n=10$, 
20, 30 and 40.  The best results on a $16^3 \times 32$ lattice with lattice spacing
$\approx 0.17$ fm, are obtained using $\alpha=0.7$ and $n=10$ for
the unimproved Wilson gauge action.  This corresponds to a
transverse RMS radius for the smeared links of 0.31 fm, where
the transverse RMS radius after $n$ sweeps is defined by
\begin{equation}
\left < r^2 \right >_n = \frac{\sum_{x} x^2 \, V_n^2 (x)}{\sum_{x}\,
V_n^2 (x)} 
\end{equation}
where
$$
V_{i}(x') = \sum_{x} \left [ (1-\alpha )\delta_{x'x} +
\frac{\alpha}{4}\sum_{\mu=1}^2 
(\delta_{x',x+\mu} + \delta_{x',x-\mu} )\right ] V_{i-1}(x)  
$$
and
$$
V_0 (x) = \left\{\begin{array}{lr}
                       1 &  x=0       \\
                       0 &  x\neq0   \\
                  \end{array} \, .
          \right.
$$
Analogous to the RMS smearing radius of Jacobi Fermion Source Smearing
\cite{Allton:1993wc}, this provides a reasonable estimate of the
transverse smearing of mean field improved links ($U_{\mu}(x) \sim
1$).

Wilson loops, $W(r,t)$, or more precisely $W(x,y,z,t)$ where $r^2 =
x^2 + y^2 + z^2$, are calculated both on-axis, along the Cartesian
directions, and off-axis.  On-axis Wilson loops are those that lie,
e.g., in the $x-t$ plane only; off-axis loops begin, for example, by
first stepping into the $y$ or $z$ (or both) directions before
proceeding through the $x-t$ plane..  This provides an alternative to the 
usual method of calculating the off axis potential by building paths in 
three different directions using small elemental squares, rectangles 
or cubes and multiplying them together to form larger paths
(see, for example, Ref.~\cite{Bolder:2000un}).  These standard techniques
for calculating the off-axis potential may be combined with the
approach described here using the variational method described extensively 
in Ref.~\cite{Morningstar:1997ff}.  

Due to the periodicity of the
lattice, the size of our Wilson loops are limited from 1 to a little
over half the smallest lattice dimension in the on-axis directions and
between 0 and 3 in the transverse directions.  For example, for a
$16^3 \times 32$ lattice, the sizes of the Wilson Loops, $t\times
x\times y\times z$, vary from a $1\times 1\times 0\times 0$ loop, to a
$10\times 10\times 3\times 3$ loop. Statistics are improved by
transposing the loops over all points on the lattice and by rotating
through the three spatial directions

In order to efficiently calculate Wilson loops of various sizes,
including off-axis loops, we build products of links in each direction
that we are considering for our Wilson loop.  Link products extending
from every lattice site are calculated in parallel.  The loops are
then formed in parallel by multiplying the appropriate sides together.
Each side is created by reusing the components of the previous loop
and one additional link.  The same approach can be extended to loops
that travel off-axis.

In an attempt to isolate the ground state potential for off-axis
paths, one symmetrizes over the path of links connecting the
off-axis heavy quark propagators, exploiting the full cubic symmetry of
the lattice.  For example, two points separated by $n_x$ sites in the 
$x$-direction, $n_y$ sites in the $y$-direction and $n_z$ sites in the 
$z$-direction, may be connected by $n_x$ link products in the $x$-direction,
$n_y$ link products in the $y$-direction and $n_z$ link products in the 
$z$-direction which we denote by the triplet $xyz$.  Instead of only 
calculating off-axis paths in the specific order $xyz$, we average over spatial 
paths calculated in the order $xyz, xzy, yxz, yzx, zxy, zyx$.  We calculated 
loops using this path-symmetrized technique as well as loops using a 
non-path-symmetrized operator where the order $xyz$ alone is considered.  
The former form of operator is designed to suppress excited states
by incorporating the full hyper-cubic symmetry of the lattice, whereas the 
latter operator is susceptible to
excited state contamination.  By comparing the static quark potential for these two 
operators, one can gain qualitative information on the effect of 
excited states in the static quark potential.

\section{Numerical Simulations}
\label{action}

The tree--level ${\cal O}(a^2)$--Symanzik-improved action \cite{Symanzik:1983dc} is defined as,

\begin{equation}
S_G=\frac{5\beta}{3}\sum_{\rm{sq}}{\cal R}e{\rm{Tr}}(1-U_{\rm{sq}}(x))-\frac{\beta}{12u_{0}^2}
\sum_{\rm{rect}}{\cal R}e{\rm{Tr}}(1-U_{\rm{rect}}(x)),
\label{gaugeaction}
\end{equation}
where the operators $U_{\rm{sq}}(x)$ and $U_{\rm{rect}}(x)$ are
defined as follows,
\begin{eqnarray}
U_{\rm{sq}}(x) & = &
U_{\mu}(x)U_{\nu}(x+\hat{\mu})U^{\dagger}_{\mu}(x+\hat{\nu})
U^\dagger_{\nu}(x)\\ 
U_{\rm{rect}}(x) & = &
U_{\mu}(x)U_{\nu}(x+\hat{\mu})U_{\nu}(x+\hat{\nu}
+\hat{\mu})
U^{\dagger}_{\mu}(x+2\hat{\nu})U^{\dagger}_{\nu}(x+\hat{\nu})
U^\dagger_{\nu}(x) \nonumber \\
& + & U_{\mu}(x)U_{\mu}(x+\hat{\mu})U_{\nu}(x+2\hat{\mu})
U^{\dagger}_{\mu}(x+\hat{\mu}+\hat{\nu})
U^{\dagger}_{\mu}(x+\hat{\nu})U^\dagger_{\nu}(x).
\label{actioneqn}
\end{eqnarray}
The link product $U_{\rm{rect}}(x)$ denotes the rectangular $1\times2$
and $2\times1$ plaquettes.  $u_0$ is the tadpole improvement factor
that largely corrects for the quantum renormalisation of the
operators. We employ the plaquette measure
\begin{equation}
u_0=\left(\frac{1}{3}{\cal R}e{\rm{Tr}}\langle U_{\rm{sq}}\rangle \right)^{1/4}.
\end{equation}
Eq.(\ref{gaugeaction}) reproduces the continuum action as
$a\rightarrow{0}$, provided that $\beta$ takes the standard
value of $6/g^2$. Perturbative corrections to this action
are estimated to be of the order of two to three
percent~\cite{Alford:1995}.

Gauge configurations are generated using the
Cabibbo-Marinari~\cite{Cab82} pseudoheat-bath algorithm with three
diagonal $SU(2)$ subgroups. Simulations are performed using a parallel
algorithm with appropriate link partitioning \cite{Bonnet:2000db}.  Configurations
are generated on a $16^3\times{32}$ lattice at  
$\beta=5.74$ using a standard Wilson action which corresponds to lattice a
spacing  $a = 0.165 (2)$, and on a 
$16^3\times{32}$ lattice at $\beta=4.38$ using
the Symanzik improved action (Eq.~\ref{actioneqn}) which also corresponds to a
lattice spacing $a = 0.165 (2)$ fm.  Thus the two lattices have
the same lattice spacing and physical volume.

Configurations are selected after 5000 thermalization sweeps from a
cold start. The mean link, $u_0$, is averaged every 10 sweeps and
updated during thermalization. For both the standard Wilson action
and the Symanzik improved action, configurations are selected every 500 sweeps. The 
following analysis is based on an ensemble of 100 configurations for each action.

\section{Simulation Results}
\label{simRes}

\begin{figure}
\begin{center}
\epsfysize=11.6truecm
\leavevmode
\rotate[l]{\epsfbox{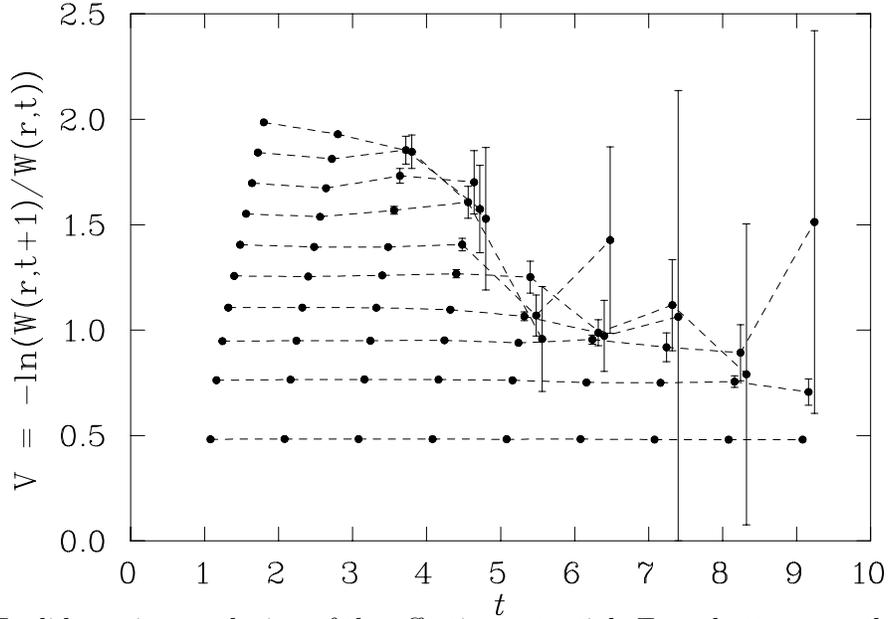}}
\caption{ Euclidean time evolution of the effective potential.  From
bottom up, the ``horizontal'' lines correspond to $r=1$ through 10.
Here we use the Symanzik improved action with $\beta = 4.38$ and 10 sweeps of 
smearing at $\alpha=0.7$.  Axes are in lattice units.  
Points are offset to the right for clarity.  }
\label{impb438n10a7Tres}
\end{center}
\end{figure}

The effective potential is obtained from
\begin{equation}
V_{t}(r) = \log\left( \frac{W(r,t)}{W(r,t+1)}\right),
\label{effpot}
\end{equation}
which is expected to be independent of $t$ for $t\gg 0$.
Figure~\ref{impb438n10a7Tres} displays the effective potential as a
function of Euclidean time, $t$, obtained from 100 configurations
generated via the Symanzik improved action at $\beta = 4.38$.  For $r\ge 7$, we
find that the signal is generally dominated by noise for $t > 4$, so
we set the upper limit of our fitting range to $t_{\rm{max}}=4$.  The
good plateau behavior at small Euclidean time is a reflection of the
optimized smearing.  Choosing the lower limit $t_{\rm{min}}=1$
leads to large $\chi^2$/d.o.f for large $r$.  We fix the fitting range to
be, in most cases, $t=2$ to 4.  The string tension is then extracted
from the ansatz,
\begin{equation}
V(r) = V_{0} + \sigma r - e/r
\label{ansatz}
\end{equation}
where $e=\pi/12$\cite{MLu}, and $V_{0}$ and $\sigma$ are fit
parameters.

The values for $V(r) + e/r = \sigma r + V_{0}$ are then fitted to a 
line and the slope, $\sigma$, and intercept, $V_{0}$, are extracted. 
We fit the range $3<r<7$ such that we are not sensitive to the coulomb
term and its associated dicretisation artifacts \cite{Edwards:1998xf}.
The error analysis is done using the Bootstrap method and all errors quoted are
statistical only.  Using
$\sqrt{\sigma}=440$ MeV to set the scale, the lattice spacings are
determined.

\begin{figure}
\begin{center}
\epsfysize=11.6truecm
\leavevmode
\rotate[l]{\epsfbox{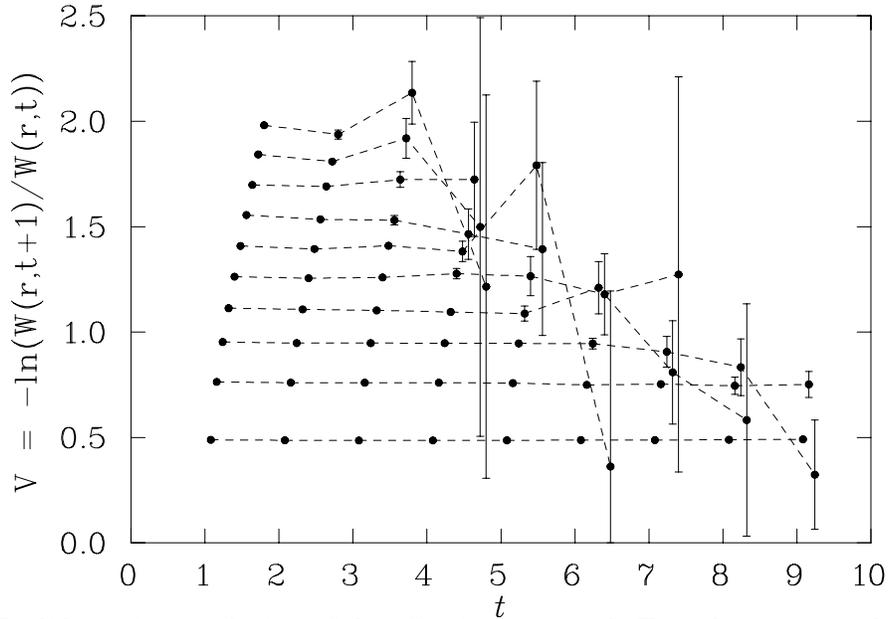}}
\caption{ Euclidean time evolution of the effective potential. From
bottom up, the ``horizontal'' lines correspond to $r=1$ through
10. Here we use the standard unimproved Wilson action with $\beta = 5.74$ and
10 sweeps of smearing at $\alpha=0.7$.  
Axes are in lattice units.  Points are offset to the right for clarity. }
\label{impb574n10a7Tres}
\end{center}
\end{figure}

\subsection{Wilson Loop Correlation Function}

To have any hope of seeing string breaking in QCD with dynamical fermions, the gauge 
links must first
be smeared to improve the overlap of the ground state. The effects of
smearing the spatial links (Eq.~(\ref{smear})) before calculating Wilson Loops
is well known.  Smearing provides access to the static quark potential at larger
distances, providing a better chance of eventually being able to
detect string breaking.  Also, the lines with smeared links exhibit
better plateau behavior in $V_t (r)$ than the unsmeared ones, indicating that we
have better isolation of the ground state.

Figs.~\ref{impb438n10a7Tres} and \ref{impb574n10a7Tres} illustrate the time dependence
of Eq.~\ref{effpot} for Symanzik improved and standard Wilson gluon actions respectively.
For clarity, we only plot values for $r$ that are obtained from on-axis loops.  For both 
the improved and unimproved actions, we see a clean signal up to $r\approx 8$
for the first three or four time slices.  For $r$ values grater than this, the quality
of the statistical signal does not allow values of $t$ larger than 2 or 3 to be included
in the fits.

\subsection{Off Axis Perturbations and Symmetry}

\begin{figure}
\begin{center}
\epsfysize=11.6truecm
\leavevmode
\rotate[l]{\epsfbox{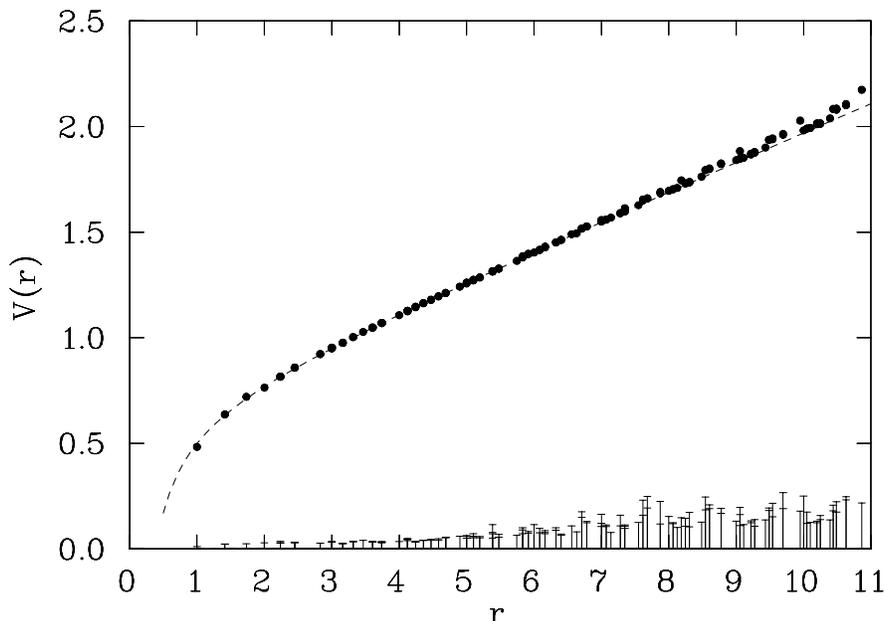}}
\caption{ The static quark potential, $V(r)$, as a function of the
separation $r$. Data is from the Symanzik improved action at $\beta = 4.38$
with 10 sweeps of smearing at $\alpha=0.7$ and a path-symmetrized operator.  Time slices 
$t=1$ to 4 are used in the fit of the correlation function. Error bars are magnified by a 
factor of 200 and placed on the $r$-axis for clarity.}
\label{impb438n10a7T14c2}
\end{center}
\end{figure}

\begin{figure}
\begin{center}
\epsfysize=11.6truecm
\leavevmode
\rotate[l]{\epsfbox{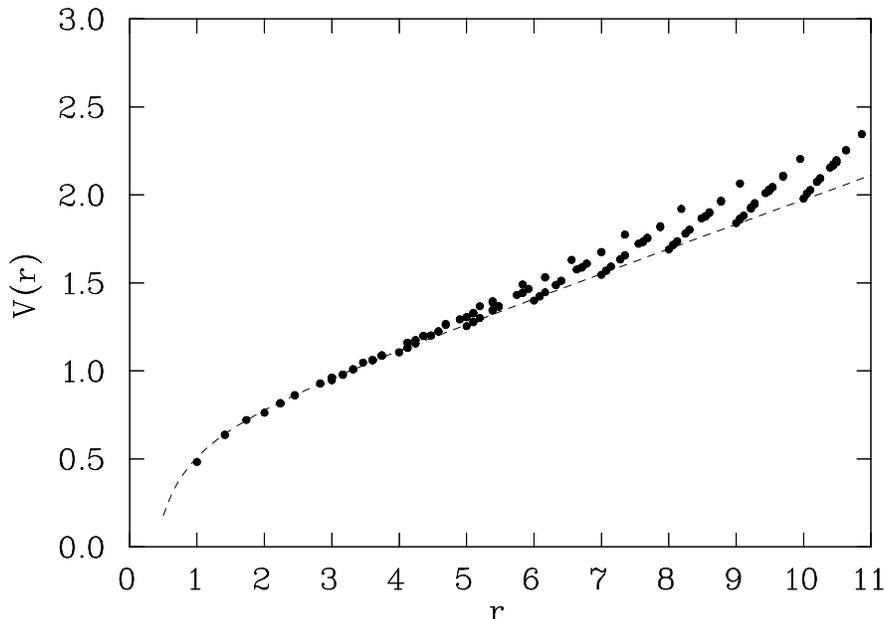}}
\caption{ The static quark potential, $V(r)$, as a function of the
separation $r$. Data is from the Symanzik improved action at $\beta = 4.38$
with 10 sweeps of smearing at $\alpha=0.7$ and a non-path-symmetrized operator.  Time slices 
$t=1$ to 4 are used in the fit of the correlation function. Statistical error bars are too small 
to be seen. }
\label{b438n10a7T12}
\end{center}
\end{figure}

Our method for stepping off-axis requires the use of a path-symmetrized
operator.  Fig.~\ref{impb438n10a7T14c2} displays results for the path-symmetrized 
operator for separating the $q\bar{q}$ pair, while Fig.~\ref{b438n10a7T12}
illustrates the non-path-symmetrized result.  Both plots are obtained with 10 smearing 
sweeps at $\alpha=0.7$.  Banding is clearly evident in the non-path-symmetrized 
case but as soon as we path-symmetrize our operators, the banding largely disappears in
off-axis points up to about $r=7$.  After $r=7$, the banding still persists
in off-axis points where loops of the form $r\times1\times3$, $r\times2\times3$ or $r\times3\times3$ 
(where $r =$ on-axis step size and is greater than 7) are used.  However the banding is negligible 
in off-axis points up to, and including, $r\times2\times2$.  Since our off-axis points are
obtained by firstly stepping in one direction and then stepping in a
different Cartesian direction, we are forming a right angle, not an
approximate straight line.  The fact that for $r<7$, the $r\times3\times3$ points lie on
the line-of-best-fit is an interesting result and provides support for the use of this
operator in a variational approach when searching for ground state dominance and its associated
string breaking in full QCD.  However, the persistence of banding at large $r$ is an indicator 
that further Euclidean time evolution is required to isolate the ground state potential.

\subsection{Effect Of Using Symanzik Improved Gauge Field Configurations}

\begin{figure}
\begin{center}
\epsfysize=11.6truecm
\leavevmode
\rotate[l]{\epsfbox{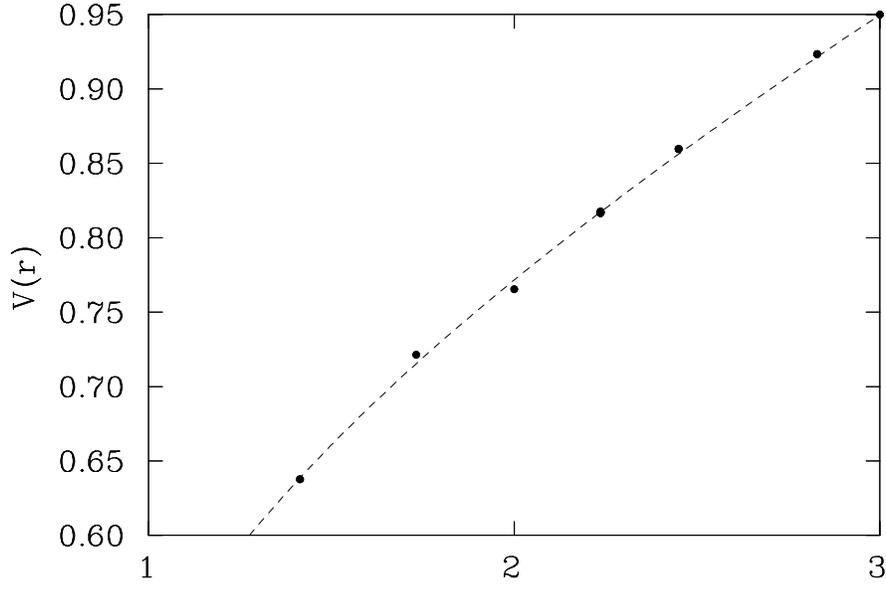}}
\caption{ A close up of the static quark potential, $V(r)$, as a function of the
separation $r$. Data is from the Symanzik improved action at $\beta = 4.38$
with 10 sweeps of smearing at $\alpha=0.7$ and a path-symmetrized operator.  Time slices 
$t=2$ to 4 are used in the fit of the correlation function. The dashed line is a fit 
of Eq.~\ref{ansatz} to on axis points $r=3$ to 7.}
\label{impb438n10a7T24zoom}
\end{center}
\end{figure}

\begin{figure}
\begin{center}
\epsfysize=11.6truecm
\leavevmode
\rotate[l]{\epsfbox{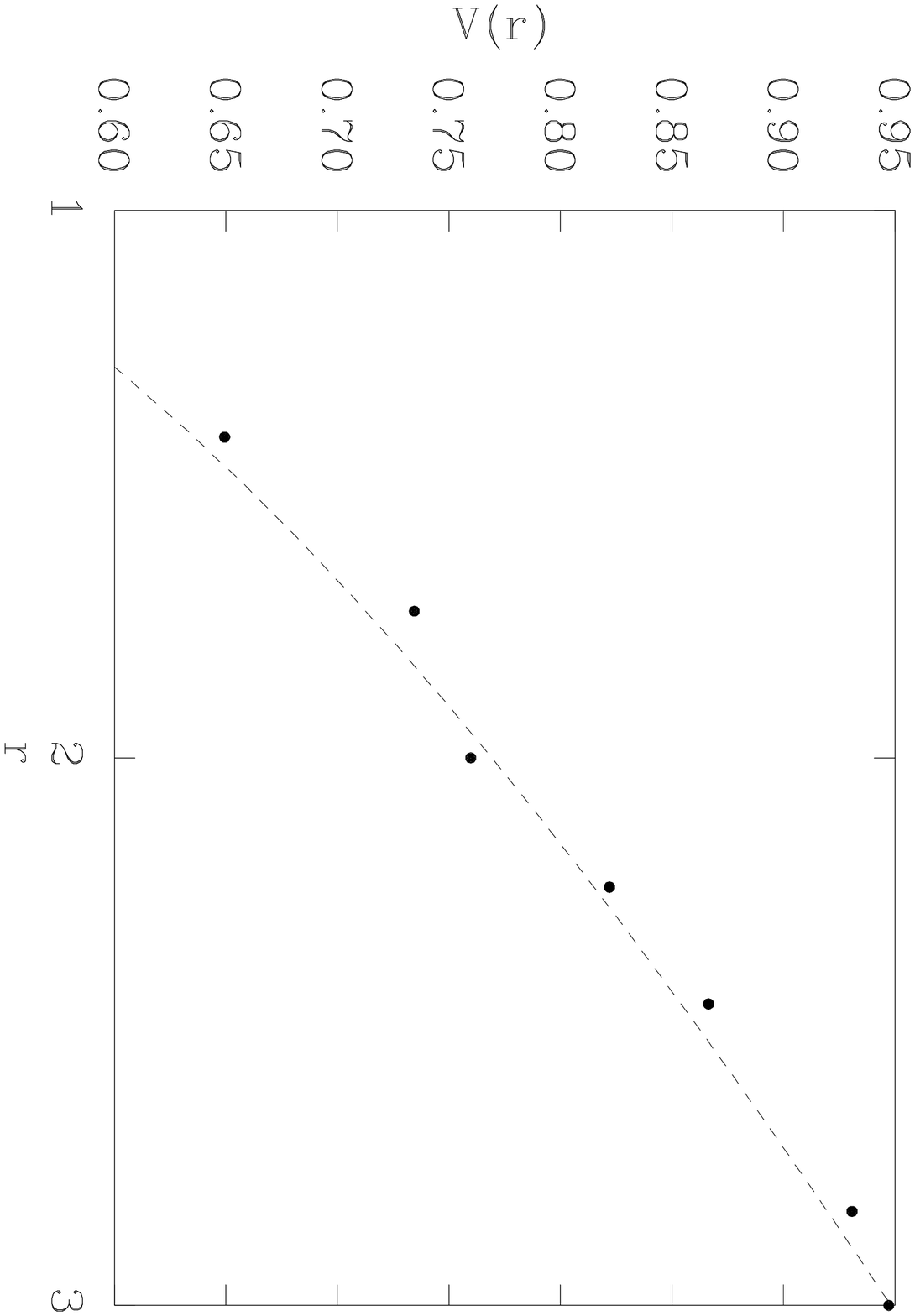}}
\caption{ A close up of the static quark potential, $V(r)$, as a function of the
separation $r$. Data is from the unimproved Wilson action at $\beta = 5.74$
with 10 sweeps of smearing at $\alpha=0.7$ and a path-symmetrized operator.  Time slices 
$t=2$ to 4 are used in the fit of the correlation function. The dashed line is a fit 
of Eq.~\ref{ansatz} to on axis points $r=3$ to 7.}
\label{impb574n10a7T24zoom}
\end{center}
\end{figure}

Historically, the main feature of improved gauge field actions is the
improved rotational symmetry \cite{Alford:1995}.  This can be seen in our 
configurations by comparing Figs.~\ref{impb438n10a7T24zoom} and \ref{impb574n10a7T24zoom}.  
These graphs are enlargements of the small-$r$ area.  The off axis points for the Symanzik improved 
$\beta=4.38$ lattices lie closer to the line of best fit through the Cartesian points, $r=$ 3 to 7,
than the unimproved Wilson $\beta=5.74$ lattices.

However, here we are most interested in the large distance properties of the Wilson loop. 
To draw firm conclusions on the merits of using Symanzik improved actions in the search for string
breaking, we create a matched set of standard Wilson gluon configurations, tuned to reproduce 
the string tension of the $\beta=4.38$ improved configurations.

\begin{figure}
\begin{center}
\epsfysize=11.6truecm
\leavevmode
\rotate[l]{\epsfbox{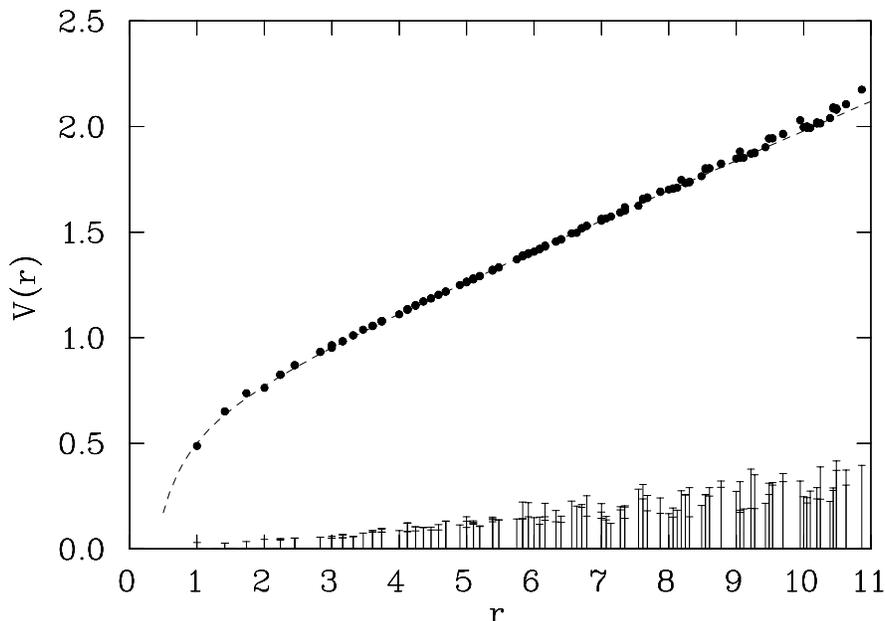}}
\caption{ The static quark potential, $V(r)$, as a function of the
separation $r$. Data is from the unimproved Wilson action at $\beta = 5.74$
with 10 sweeps of smearing at $\alpha=0.7$ and a path-symmetrized operator.  Time slices 
$t=1$ to 4 are used in the fit of the correlation function. Error bars are magnified by a
factor of 200 and placed on the $r$-axis for clarity.}
\label{impb574n10a7T14}
\end{center}
\end{figure}

\begin{figure}
\begin{center}
\epsfysize=11.6truecm
\leavevmode
\rotate[l]{\epsfbox{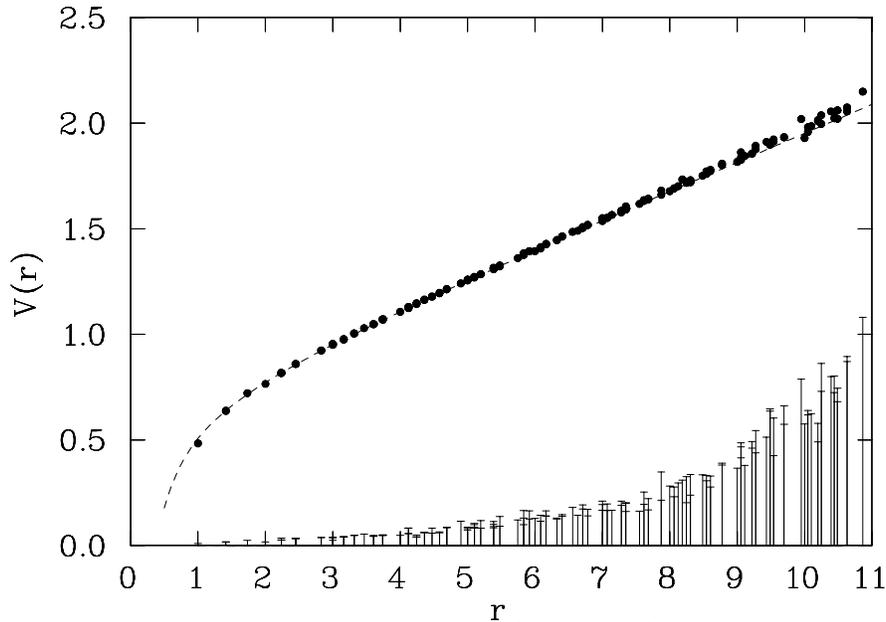}}
\caption{ The static quark potential, $V(r)$, as a function of the
separation $r$. Data is from the Symanzik improved action at $\beta = 4.38$
with 10 sweeps of smearing at $\alpha=0.7$ and a path-symmetrized operator. Here time slices
$t=2$ to 4 are used in the fit of the correlation function. Error bars are magnified by a
factor of 200 and placed on the $r$-axis for clarity.}
\label{impb438n10a7T24}
\end{center}
\end{figure}

\begin{figure}
\begin{center}
\epsfysize=11.6truecm
\leavevmode
\rotate[l]{\epsfbox{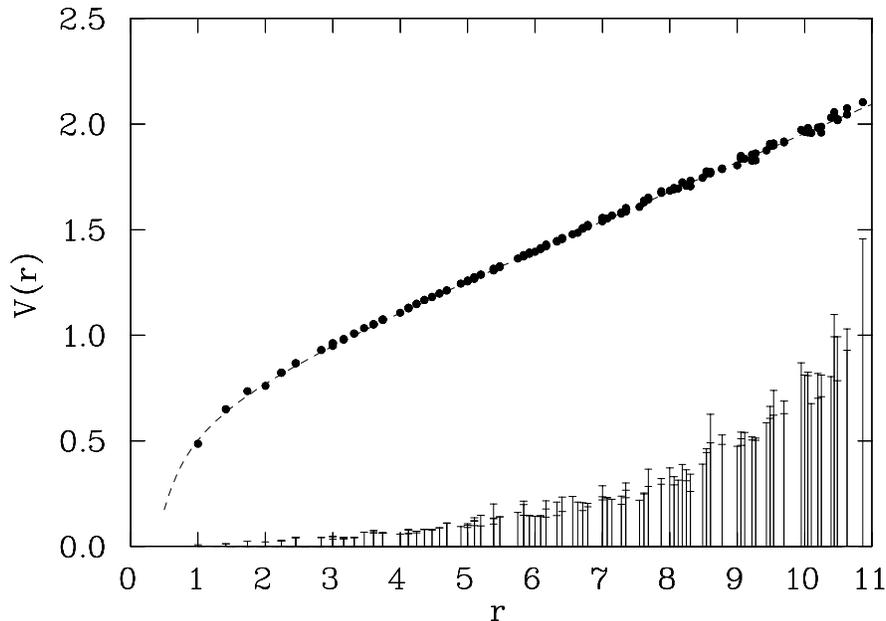}}
\caption{ The static quark potential, $V(r)$, as a function of the
separation $r$. Data is from the unimproved Wilson action at $\beta = 5.74$
with 10 sweeps of smearing at $\alpha=0.7$ and a path-symmetrized operator.  Time slices 
$t=2$ to 4 are used in the fit of the correlation function. Error bars are magnified by a
factor of 200 and placed on the $r$-axis for clarity.}
\label{impb574n10a7T24}
\end{center}
\end{figure}

Figs.~\ref{impb438n10a7T14c2} and \ref{impb574n10a7T14} compare Symanzik improved and unimproved 
Wilson results for $V(r)$ extracted from correlated fits within the range $1\le t\le 4$.  A similar 
comparison is made with Figs.~\ref{impb438n10a7T24} and \ref{impb574n10a7T24} within the range 
$2\le t\le 4$.  Magnified error bars are plotted on the X-axis to allow a comparison of signal to 
noise between the two actions.  Exponential growth in the error bars for large $r$ is apparent in 
the fits with $2\le t\le 4$.

It is crucial to compare matched lattices with the same lattice spacing and same physical 
volume.  For example, a comparison of an unimproved Wilson lattice with $\beta=5.70$ and lattice 
spacing $a=0.181(2)$, with a Symanzik improved lattice at $\beta=4.38$ and $a=0.165(2)$ leads one 
to incorrectly conclude that Symanzik improved actions lead to an improvement in the 
signal-to-noise ratio of the static quark potential at large $q\bar{q}$ separations.  The 
unimproved configurations with $\beta=5.70$ lose the signal around $r/a=7.5$ ($r=1.37$ fm) when 
using time slices $t=2-4$, whereas the improved configurations hold the signal up to $r/a=9.5$ 
($r=1.59$ fm) when using $t=2-4$.  This effect is due to the slightly larger lattice spacing in 
the $\beta=5.70$ simulations, which spoils the signal-to-noise ratio even after the larger lattice
spacing is taken into account.

Close inspection of Figs.~\ref{impb574n10a7T14} and \ref{impb574n10a7T24}, where the Euclidean
time regimes $1\le t\le 4$ and $2\le t\le 4$ are compared, reveals how the off-axis points can 
also be used to gain confidence in ground state dominance.  In Fig.~\ref{impb574n10a7T14},
banding is apparent.  However, upon adjusting the Euclidean time regime for the fit to larger
times in Fig.~\ref{impb574n10a7T24}, the banding is largely removed.

\section{Conclusions}
\label{conclusion}

We have calculated the static quark potential in quenched QCD using
Symanzik improved and unimproved Wilson gluon actions.  We have kept the lattice
spacing and the physical volume of these lattices equal
so that we can meaningfully compare the results.  The number of gauge field configurations (100 here)
is also held fixed for each action.
We have explicitly shown that, despite the relatively coarse lattice spacing, the unimproved and 
computationally less expensive Wilson action does just as well as the improved action in extracting the 
$q\bar{q}$ potential at large separations.  If one wishes to keep non-perturbative physics such 
as non-trivial topological fluctuations on the lattice, then one needs $a<0.15$ fm \cite{ImpSmear}, and thus
$r/a >7$.  In this case, the unimproved, standard Wilson gauge action is ideal for todays string 
breaking searches and computational resources can be redirected elsewhere.  Another advantage for 
using unimproved Wilson gauge configurations is that we recover the extremely useful method for 
calculating the overlap with the ground state, $C_1 (r)$, and thus tuning the smearing parameters.

We also explored the use of unconventional paths in accessing off-axis values of $r$ in the static
quark potential.  These paths can provide insight into the extent to which the ground state potential
dominates the Wilson loop at large Euclidean times.  Provided the paths are symmetrized, these new 
paths provide useful information on the ground state potential and nearby excited potentials.  Combined 
with standard paths and variational techniques, these paths offer additional promise for the search 
for string breaking in lattice QCD.

\newpage
\acknowledgments

This work was supported by the Australian Research Council and by
grants of supercomputer time on the CM-5 made available through the
South Australian Centre for Parallel Computing.  We would also like to
thank the National Computing Facility for Lattice Gauge Theory for the
use of the Orion Supercomputer.  AGW also acknowledges
support form the Department of Energy Contract No.~DE-FG05-86ER40273
and by the Florida State University Supercomputer Computations
Research Institute which is partially funded by the Department of
Energy through contract No.~DE-FC05-85ER25000

\end{document}